\documentclass[%
 reprint,
 amsmath,amssymb,
 longbibliography,
 prx
]{revtex4-2}

\usepackage[ruled]{algorithm2e}
\usepackage[noend]{algpseudocode}
\usepackage{wasysym}
\usepackage{graphicx}
\usepackage{dcolumn}
\usepackage{bm}
\DeclareMathAlphabet{\mathpzc}{OT1}{pzc}{m}{it}

\begin{document}

\preprint{APS/123-QED}

\title{A simple framework for contrastive learning phases of matter}

\author{Xiao-Qi Han}
\affiliation{Department of Physics, Renmin University of China, Beijing 100872, China}

\author{Sheng-Song Xu}
\affiliation{Department of Physics, Renmin University of China, Beijing 100872, China}

\author{Zhen Feng}
\affiliation{Department of Physics, Renmin University of China, Beijing 100872, China}

\author{Rong-Qiang He}\email{rqhe@ruc.edu.cn}
\affiliation{Department of Physics, Renmin University of China, Beijing 100872, China}

\author{Zhong-Yi Lu}\email{zlu@ruc.edu.cn}
\affiliation{Department of Physics, Renmin University of China, Beijing 100872, China}

\date{\today}

\begin{abstract}
  A main task in condensed-matter physics is to recognize, classify, and characterize phases of matter and the corresponding phase transitions, for which machine learning provides a new class of research tools due to the remarkable development in computing power and algorithms. Despite much exploration in this new field, usually different methods and techniques are needed for different scenarios. Here, we present {\em SimCLP}: a simple framework for contrastive learning phases of matter, which is inspired by the recent development in contrastive learning of visual representations. We demonstrate the success of this framework on several representative systems, including classical and quantum, single-particle and many-body, conventional and topological. SimCLP is flexible and free of usual burdens such as manual feature engineering and prior knowledge. The only prerequisite is to prepare enough state configurations. Furthermore, it can generate representation vectors and labels and hence help tackle other problems. SimCLP therefore paves an alternative way to the development of a generic tool for identifying unexplored phase transitions.
\end{abstract}


\maketitle


\section{Introduction}

The problem of classifying phases of matter has lasted for centuries and more and more states of matter have been discovered. \cite{sachdev2011QPT} In recent years, various types of machine learning methods have been applied on this subject \cite{Carleo2019RMP,dunjko2018machine,Carrasquilla2021How,Carrasquilla2020review,Uvarov2020processor,Bai2021compression,Bohrdt2020Analyzing}. According to whether or not the data need to be labeled, they are mainly divided into two categories: supervised and unsupervised methods. The data usually consist of state configurations (e.g. samples from a Monte Carlo) or some other information deliberately prepared (e.g. entanglement spectra derived from wave functions), which serve as input to machine learning. The labels are usually our target, namely to which phases the data belong. Supervised methods can indeed learn phases efficiently when the labels are available, \cite{Juan2017machine,Kelvin2017Fermions,Beach2018vortices,Venderley2018Equilibrium,Deng2017MachineTopo,sancho2021kernels, Driskell2021NFL,havlivcek2019supervised,Schindler2017MBL,Zhang2019percolation,Zhai2018Topological,Yusuke2021machine,Hugo2019MBL} but their applications are limited since in most cases the labels are unavailable. In contrast, unsupervised methods do not require labels. They recognize phases by extracting features or clustering the data. Some unsupervised methods, such as principal component analysis (PCA) and variational autoencoder (VAE), are easy to implement and work well for simple systems (e.g. the two-dimensional Ising model), but fail for complex systems. \cite{Nieuwenburg2017learning,Wang2016unsupervised,Kharkov2020chaos,Wetzel2017Unsupervised,Hu2017PCA} Some other unsupervised methods are tricky and technically difficult to implement. \cite{Nieuwenburg2017learning,rodriguez2019unsupervised,Huembeli2018adversarial,broecker2017quantum,Balabanov2020topological,Canabarro2019Unveiling, greplova2020unsupervised,Wang2021index,Wang2020Nuclear,Che2020Topological,Shen2021Supervised,Ni2019complex,Lee2019Confusion}

Inspired by recent progress in contrastive learning of visual representations, \cite{2020SimCLR,wang2022molecular,Deng2021CReSS} in this paper we propose a simple framework for constrastive learning phases of matter (dubbed as {SimCLP}). It contains two identical neural networks (with same architecture and parameters) and does not require labels. The outputs of one of them serve as labels for the other and vice verse. Therefore, SimCLP is unsupervised but its two neural networks can be trained like doing a supervised machine learning. In such a way, SimCLP combines the merits of both supervised and unsupervised methods and circumvents their drawbacks.

Furthermore, the output of each neural network is a representation vector to the input. The input data from physical systems with the same conditions should be physically similar, hence the corresponding representation vectors should also be similar. Therefore, the training target is to maximize the similarity between these representation vectors. After the two neural networks are fully trained, we can readily predict phases and their transitions by quantifying the similarity between representation vectors for input data from physical systems with different conditions.

We would like to emphasize a number of points. Our framework is flexible. The architecture of the involved neural networks is not restricted, and various excellent neural networks in the AI area can be adopted straightforwardly. {SimCLP} does not require any prior knowledge, such as data labels, Hamiltonians, order parameters, how many phases are involved, etc. Preparing enough training data (i.e. state configurations) is the only prerequisite and is key to avoid overfitting in the training. As valuable by-products, {SimCLP} can generate representation vectors and labels and help tackle other problems. For example, they can be utilized to set up other supervised learning for other purposes. \cite{Kelvin2017Fermions,Venderley2018Equilibrium,Bai2021compression,Hugo2019MBL,Liu2020Random}

We demonstrate our framework and practical implementation with several representative model systems: (1) the two-dimensional Ising model, which is a classical system developing a long-range magnetic order accompanied by spontaneous symmetry breaking below a certain temperature; (2) the quantum compass model, which is a quantum many-body system spontaneously breaking a directional symmetry below a certain temperature; (3) the Aubry-Andr\'e model, which is a disordered fermion system with quasi-periodic potentials experiencing a metallic-to-insulating Anderson transition; (4) the Su-Schrieffer-Heeger (SSH) model, which features a topological phase transition protected by chiral symmetry. We predict correctly all the phases without using any prior knowledge. The only prerequisite is to prepare enough state configurations as input data to the neural networks, which are routinely generated with standard Monte Carlo simulations.

\section{The {\em SimCLP} framework}

\begin{figure}[htp!]
  \includegraphics[width=6cm]{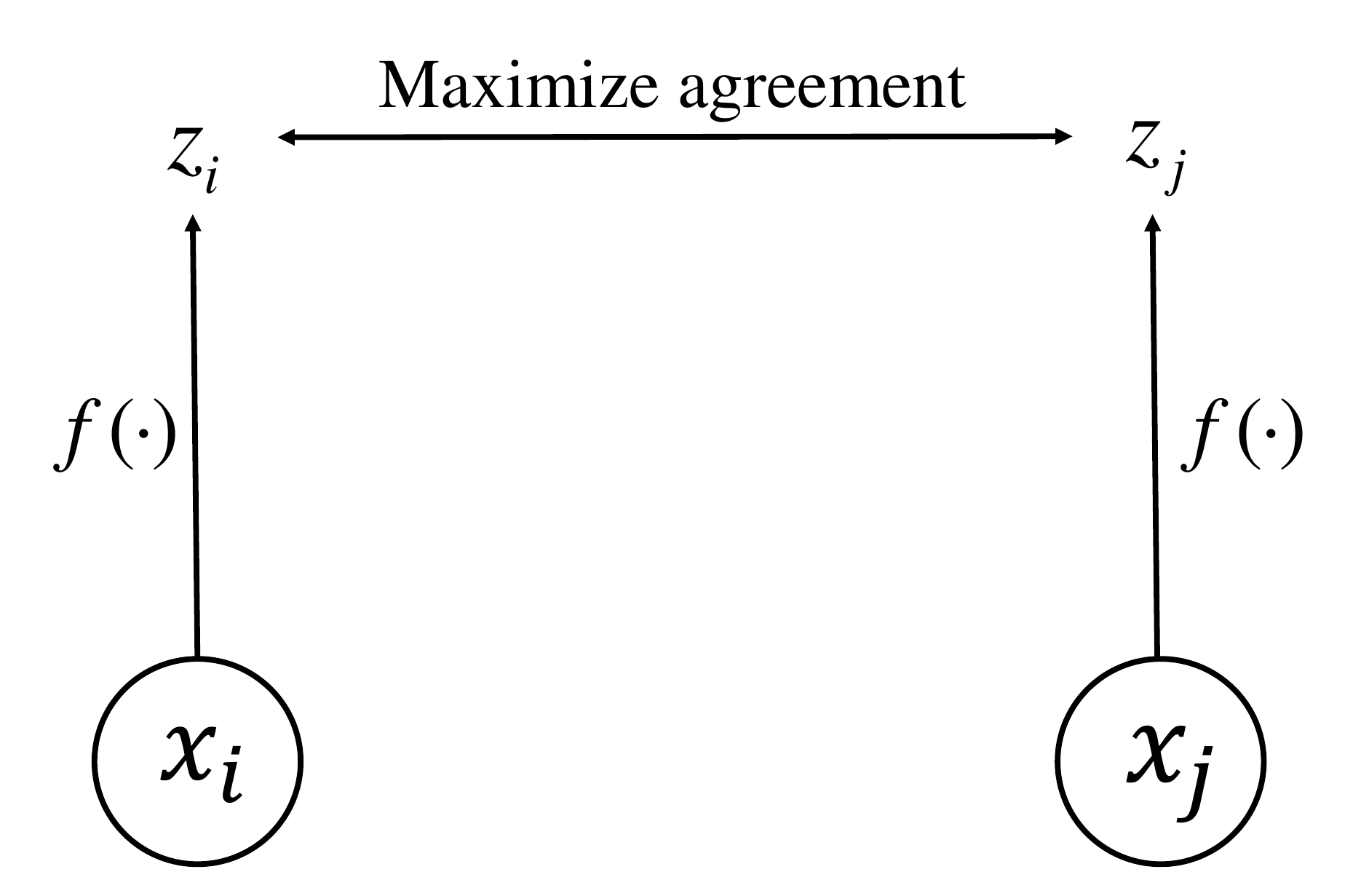}
  \caption{A framework for contrastive learning of phases of matter. Two samples ${x_i}$ and ${x_j}$ are drawn from a training data set, which are a positive pair if drawn from the data set with the same model parameter $\eta$, otherwise a negative pair. An encoder network $f(\cdot)$ is trained to maximize agreement using a contrastive loss of the two corresponding representation vectors $z_i$ and $z_j$.}
  \label{fig:net}
\end{figure}

Inspired by the recent contrastive learning algorithm SimCLR \cite{2020SimCLR}, our SimCLP framework learns phases of matter by maximizing agreement between two samples from a training data set with the same model parameter $\eta$ via a contrastive loss of the two corresponding representation vectors. As illustrated in  Fig.~\ref{fig:net}, this framework comprises the following four major components.
\begin{itemize}

  \item For a specific physical model with a variable model parameter $\eta$, which may encounter a phase transition at certain point $\eta = \eta_c$, we generate a series of training sets $\{ \Omega_k \}$ and testing sets $\{ \Theta_k \}$ (e.g. samples of state configurations drawn from a Monte Carlo simulation) with model parameter $\eta = \eta_k$, $k = 1, 2, \ldots, N$. To avoid overfitting in the contrastive learning, each training set has to be large enough. The best performance will be achieved when every sample in each training set is used only once. This is not a hard task as a Monte Carlo algorithm usually can generate samples easily.

  \item Two samples ${x_i}$ and ${x_j}$ are drawn from the training data sets, which are a positive pair if drawn from the data set with the same model parameter $\eta = \eta_k$, otherwise a negative pair.

  \item A neural network encoder $f(\cdot)$ that extracts representation vectors ($z_i$ and $z_j$) from data examples ($x_i$ and $x_j$). Our framework allows various choices of the network architecture without any restriction. For simplicity we adopt multilayer perceptron (MLP) and LeNet \cite{LeCun1989LeNet} to obtain $z_i = f(x_i) = \mathrm{MLP}(x_i)$ or $\mathrm{LeNet}(x_i)$, where $z_i \in \mathbb{R}^d$ is the output of the neural network. A contrastive loss is defined on $z_i$'s.

  \item Following SimCLR \cite{2020SimCLR}, a contrastive loss function is defined for a contrastive prediction task. Given a set $\{ x_l \}$ ($l = 1, 2, \ldots, 2N$ for example) including a positive pair of examples $x_i$ and $x_j$, the contrastive prediction task aims to identify $x_j$ in $\{ x_l \}_{l \ne i}$ for a given $x_i$.

\end{itemize}

We randomly sample two examples from each traing set with model parameter $\eta = \eta_k$, resulting in a minibatch of $2N$ data points, on which a contrastive prediction task is defined. Similar to SimCLR \cite{2020SimCLR}, we do not sample negative examples explicitly. Instead, given a positive pair, we treat the other $2(N - 1)$ examples within a minibatch as negative examples. The cosine similarity is defined for two vectors $u$ and $v$ as $\mathrm{sim}(u,v)={{u}^{T}}v/\left\| u \right\|\left\| v \right\|$, which is the dot product between ${{\ell }_{2}}$-normalized $u$ and $v$. Then the loss function for a positive pair of examples $(i,j)$ is defined as
\begin{equation}
  {{\ell }_{i,j}}=-\log \frac{\exp (\mathrm{sim}({{z}_{i}},{{z}_{j}})/\tau )}{\underset{l=1,l \ne i}{\overset{2N}{\mathop \sum }}\,\exp (\mathrm{sim}({{z}_{i}},{{z}_{l}})/\tau )},
  \label{eq:loss}
\end{equation}
where $\tau$ denotes a temperature parameter for the neural network, which should not be confused with the temperature $T$ for a physical system. The final loss (the normalized temperature-scaled cross entropy loss (NT-Xent)) is computed across all positive pairs, both $(i,j)$ and $(j,i)$, in a mini-batch. Algorithm~\ref{alg:CL} summarizes the proposed framework.

\begin{algorithm}[h]
\caption{SimCLP's learning algorithm}
\label{alg:CL}

{\bf preparation:} generate a series of training sets $\{ \Omega_k \}$ and testing sets $\{ \Theta_k \}$ (e.g. samples of state configurations drawn from a Monte Carlo simulation) for a specific physical model with model parameter $\eta = \eta_k$, $k = 1, 2, \ldots, N$.

$\textbf{input}$: batch size $N$, constant $\tau$, structure of $f$

$\textbf{for}$ sampled minibatch $\{{{{x}}_{k}}\}_{k=1}^{N}$ $\textbf{do}$

~~~~ $\textbf{for all}$ $k \in \{1, \ldots, N\}$ $\textbf{do}$

~~~~~~~~ draw two samples $x_{2k-1}$ and $x_{2k}$ from \

~~~~~~~~ the training data set with model parameter $\eta = \eta_k$

~~~~ $\textbf{end for}$

~~~~ $\textbf{for all}$ $i \in \{1, \ldots, 2N\}$ and $j \in \{1, \ldots, 2N\}$ $\textbf{do}$

~~~~~~~~ ${{s}_{i,j}} = \frac{z_{i}^{\top }{{z}_{j}}}{\left\| {{z}_{i}} \right\|\left\| {{z}_{j}} \right\| }$ ~~~~ \# pairwise cos similarity

~~~~ $\textbf{end for}$

~~~~ $\textbf{define}$  $\ell (i,j)=-\log \frac{\exp \left( {{s}_{i,j}}/\tau \right)}
{\sum\limits_{l=1, l \ne i}^{2N} \exp \left( {s_{i,l}}/\tau  \right)}$

~~~~ $\mathcal{L} = \frac{1}{2N} \underset{k=1}{\overset{N}{\mathop \sum }}\,[\ell(2k-1,2k) + \ell(2k,2k-1)]$

~~~~ update networks $f$ to minimize $\mathcal{L}$

$\textbf{end for}$

{\bf prediction:} quantify the similarity between two differently conditioned systems with model parameters $\eta_k$ and $\eta_{k\prime}$ by averaging the cos similarity of pair samples $x_k$ and $x_{k\prime}$ drawn from the testing sets with model parameters $\eta_k$ and $\eta_{k\prime}$ to predict phases and their transitions.

\end{algorithm}

\section{Numerical results and discussions}

\subsection{Contrastive learning phases of the Ising model}

First, we apply our method on the prototypical example of the square-lattice ferromagnetic Ising model,
\begin{equation}
  H=-J\sum\limits_{\left\langle ij \right\rangle }{\sigma _{i}^{z}\sigma _{j}^{z}},
  \label{eq:isingmodel}
\end{equation}
where the Ising variables $\sigma _{i}^{z} = \pm 1$. We set $J = 1$ as the energy unit. For a system with linear size $L$, there are $N_s = L^2$ lattice sites and hence the state
space is of size $2^{N_s}$. There is a well-understood phase transition \cite{1944Crystal} at temperature $T_c = 2.269$, separating the high-temperature disordered phase and the low-temperature ferromagnetic phase. The standard Monte Carlo method is employed to generate enough uncorrelated state configurations to constitute the training data sets and testing data sets with $N = 51$ temperatures $T_k = 1 + (k - 1) \Delta T$ and $\Delta T = 0.05$. We adopt an MLP with one hidden layer (consisting of 10 neurons) as the encoder neural network to obtain $z_i = f(x_i) = W^{(2)}\sigma(W^{(1)} x_i)$ where $\sigma$ is a ReLU nonlinearity and $W = \{W^{(1)}, W^{(2)} \}$ are learnable parameters. The Adam algorithm is used for neural network optimization. To avoid overfitting, each example in the training sets is used only once.

After the neural network is trained, we try to predict possible phase transitions by calculating the adjacent similarity and mutual similarity with the testing sets. The adjacent similarity at a temperature $T$ is the averaged cosine similarity of the representation vectors of pair examples from two testing sets with $T - \Delta T$ and $T + \Delta T$, which measures the similarity between state configurations with the two temperatures $T \pm \Delta T$. Similarly, the mutual similarity is defined for any two temperatures $T$ and $T^{\prime}$ and hence measures the similarity between state configurations with any two temperatures.

\begin{figure}[htp!]
  \includegraphics[width=8cm]{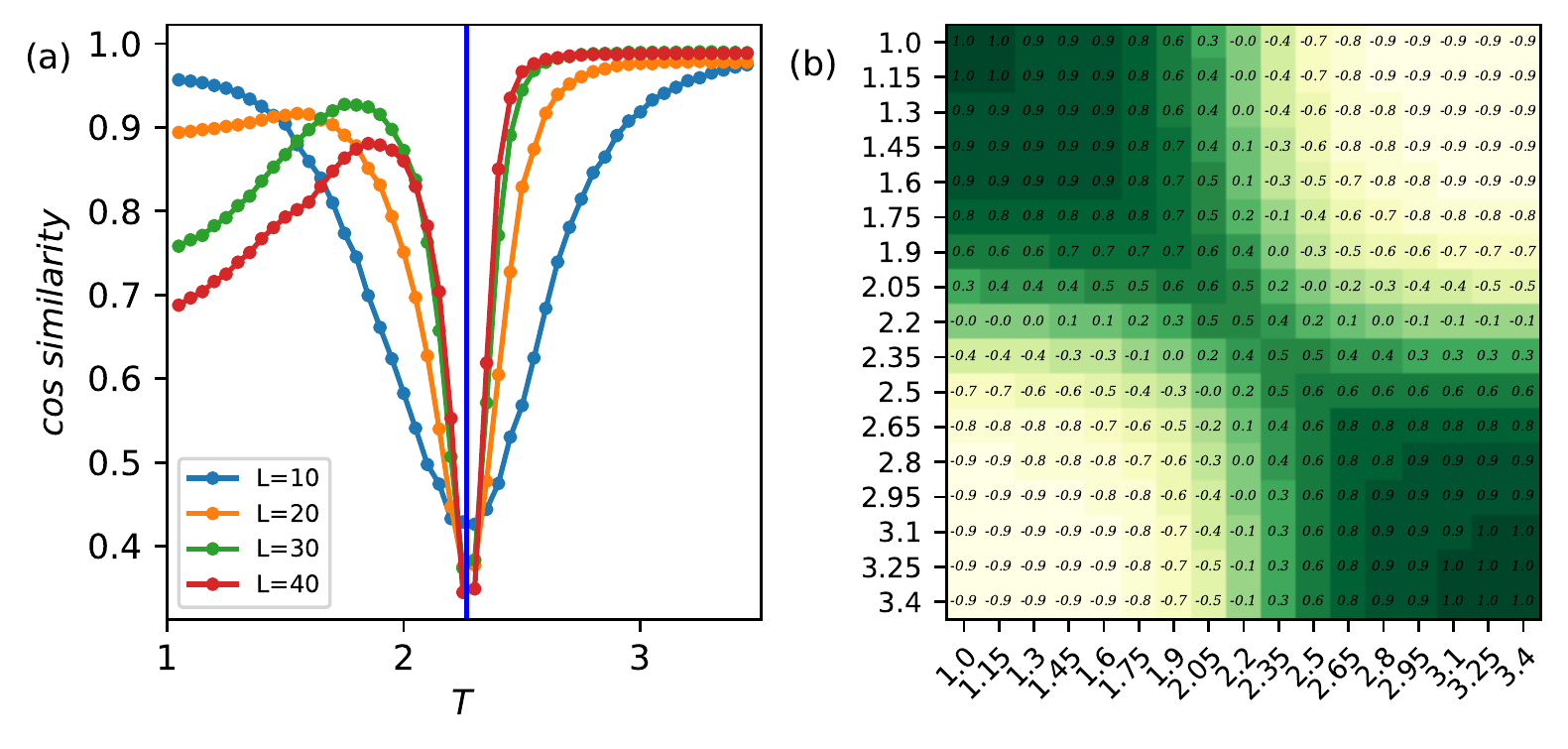}
  \caption{Cosine similarities estimated for the two-dimensional Ising model. (a) Adjacent similarity, which measures the similarity between state configurations with temperatures $T - \Delta T$ and $T + \Delta T$. The linear size of the model $L = 10, 20, 30$, and $40$. (b) Mutual similarity with $L = 10$, which measures the similarity between state configurations with any two temperatures $T$ and $T^{\prime}$.}
  \label{fig:ising}
\end{figure}

As shown in Fig.~\ref{fig:ising}(a), the adjacent similarity is close to 1 (saturation value) for temperatures far away from $T_c$ and presents a V-shape around $T \sim T_c$. The mutual similarity for two temperatures $T$ and $T^{\prime}$ decreases as $\left| T - T^{\prime} \right|$ increases. This is because physically the state configurations are similar in the same phase, but not similar (qualitatively different) in different phases. Near the phase transition point, the structure of the state configurations varies quickly and hence the adjacent similarity drops, implying a phase transition point (Fig.~\ref{fig:ising}(a)). As shown in Fig.~\ref{fig:ising}(b), the mutual similarities are high for state configurations from the same phase and low otherwise, successfully discriminating different phases of matter. In such a way, our method successfully captures this structural varying feature and leverages it to predict correctly the phases and their transition point.

\begin{figure}[htp!]
  \includegraphics[width=8cm]{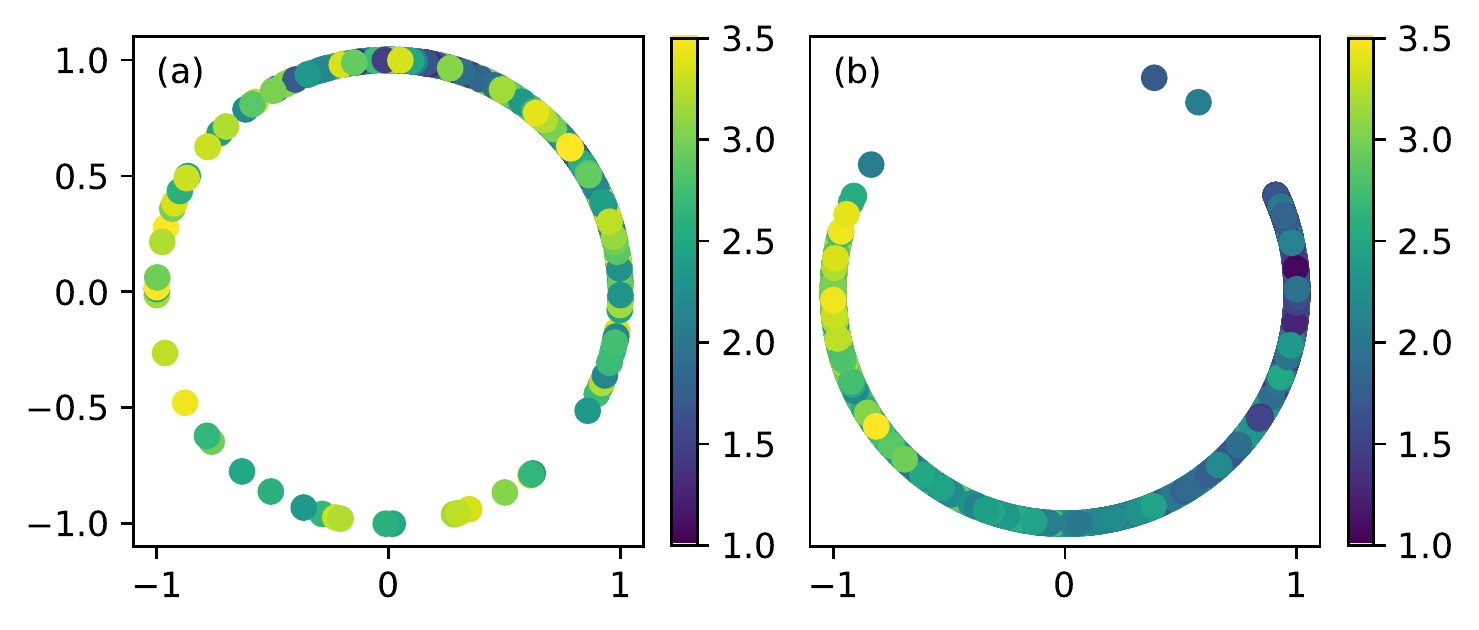}
  \caption{Representation vectors of the state configurations of the Ising model with different temperatures, which are of dimension 2 and hence are located on a circle when normalized. (a) Before training the encoder neural network, the representation vectors are located randomly on the circle. (b) After training, the representation vector moves on the circle as the temperature varies.}
  \label{fig:repvec}
\end{figure}

Further, we examine details of the representation vectors $z_i$ of the state configurations, which is visualized in Fig.~\ref{fig:repvec}. The dimension of a representation vector is not limited. For simplicity and visuality, we choose two dimensions in our demonstration and hence the representation vectors are located on a unit circle when normalized. Before training the encoder neural network, the representation vectors are located randomly on the circle, while after training, they move on the circle as the temperature varies and the high- and low-temperature ones cluster, respectively.

\subsection{Contrastive learning phases of the quantum compass model}

As above, our method works successfully for a classical physical system. Now, we test our method on a quantum system---the quantum compass model
\begin{equation}
  H = -\frac{1}{4}{{J}_{x}}\sum\limits_{j}{{{X}_{j}}}{{X}_{j+{{e}_{x}}}}-\frac{1}{4}{{J}_{z}}\sum\limits_{j}{{{Z}_{j}}}{{Z}_{j+{{e}_{z}}}},
  \label{eq:compassmodel}
\end{equation}
where $e_x$ and $e_z$ are unit vectors along $x$ and $z$ directions, respectively, and $X_j$ and $Z_j$ represent Pauli $x$ and $z$ operators at lattice site $j$, respectively. An order parameter is proposed in Ref.~\onlinecite{Mishra2004Directional}, $Q = \left| {{\left\langle {{X}_{j}}{{X}_{j+{{e}_{x}}}}-{{Z}_{j}}{{Z}_{j+{{e}_{y}}}} \right\rangle }} \right|$, which is zero when the temperature $T > T_c = 0.0585(3)$ \cite{Wenzel2010compass} and nonzero when $T < T_c$. The system is disordered and the $x$ and $z$ directions are symmetrically equivalent at high temperatures. However, at temperatures below $T_c$, the system spontaneously breaks the directional symmetry and the $x$ and $z$ directions are no longer equivalent and hence the order parameter is finite. We set $J = J_x = J_z = 1$ as the energy unit. The stochastic series expansion (SSE) quantum Monte Carlo method \cite{Sandvik2002SSEdirected} is employed to generate enough uncorrelated state configurations to constitute the training data sets and testing data sets with $N = 51$ temperatures $T_k = 0.01 + (k - 1) \Delta T$ and $\Delta T = 0.003$. Here, the LeNet \cite{LeCun1989LeNet} is used as the encoder neural network, which consists of a two-layer convolutional neural network and followed by a two-layer fully connected neural network.

\begin{figure}[htp!]
  \includegraphics[width=8cm]{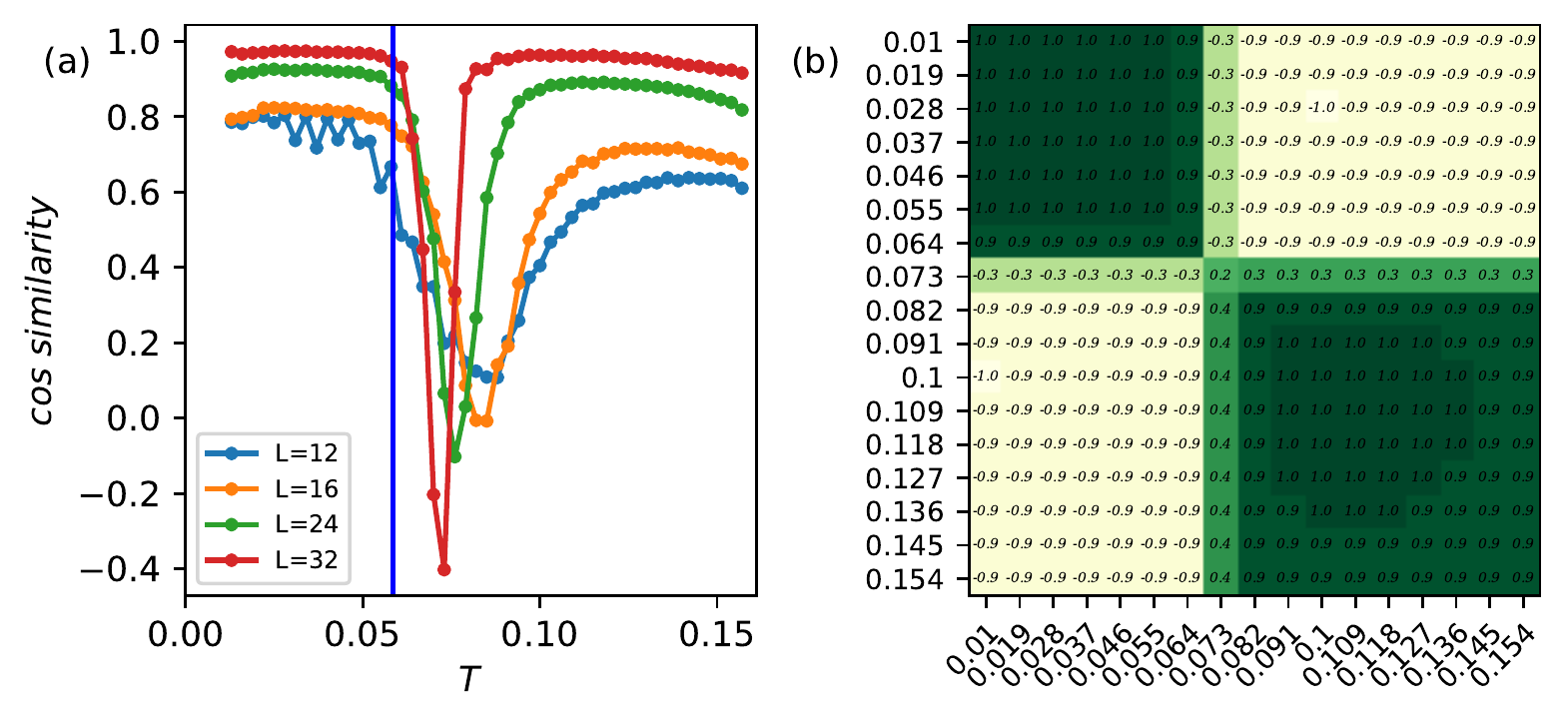}
  \caption{Cosine similarities estimated for the quantum compass model. (a) Adjacent similarity, which measures the similarity between state configurations with temperatures $T - \Delta T$ and $T + \Delta T$. The linear size of the model $L = 12, 16, 24$, and $32$. (b) Mutual similarity with $L = 32$, which measures the similarity between state configurations with any two temperatures $T$ and $T^{\prime}$.}
  \label{fig:compass}
\end{figure}

As shown in Fig.~\ref{fig:compass}, similar to the Ising model case, the adjacent similarity is close to 1 (saturation value) for temperatures far away from $T_c$ and presents a V-shape around $T \sim T_c$. As $L$ increases, the position of the dip moves towards the true transition point (noting the severe finite-size effect in this model \cite{Wenzel2010compass}). Our method successfully predicts the phase transition point and works also for this quantum many-body system.

\subsection{Contrastive learning phases of the Aubry-Andr\'e model}

In both the above examples, the phase transitions are driven by temperature. Here, we study the Aubry-Andr\'e model of spinless fermions \cite{aubry1980analyticity}, which is a quantum disordered model and experiences a quantum phase transition driven by a quasiperiodic potential $\lambda$,
\begin{equation}
  H = -J \sum\limits_{i}{(c_{i}^{+}{{c}_{i+1}}+h.c.)} + 2\lambda \sum\limits_{i} \mathrm{cos}(2\pi \phi i)c_{i}^{+}{{c}_{i}},
  \label{eq:AAmodel}
\end{equation}
where $\phi $ is the golden ratio, $\lambda $ is the amplitude of a quasiperiodic potential coupled to the fermionic density $n_i = c_i^+ c_i$, and $c_{i}^{+}$ and ${{c}_{i}}$ are standard fermionic operators. In this system the noninteracting fermions transition from a metallic to an Anderson localized phase at the critical value of the quasiperiodic potential ${{\lambda }_{c}} / J = 1$. We set $J = 1$ as the energy unit. Standard variational Monte Carlo provides us with state configurations weighted by the amplitude squared of the single-particle ground state wave function. The training data sets and testing data sets are prepared for $N = 51$ $\lambda$'s, $\lambda_k = k \Delta \lambda$ and $\Delta \lambda = 0.04$. The two-layer MLP is used as the encoder neural network, the same as the Ising model case above.

\begin{figure}[htp!]
  \includegraphics[width=8cm]{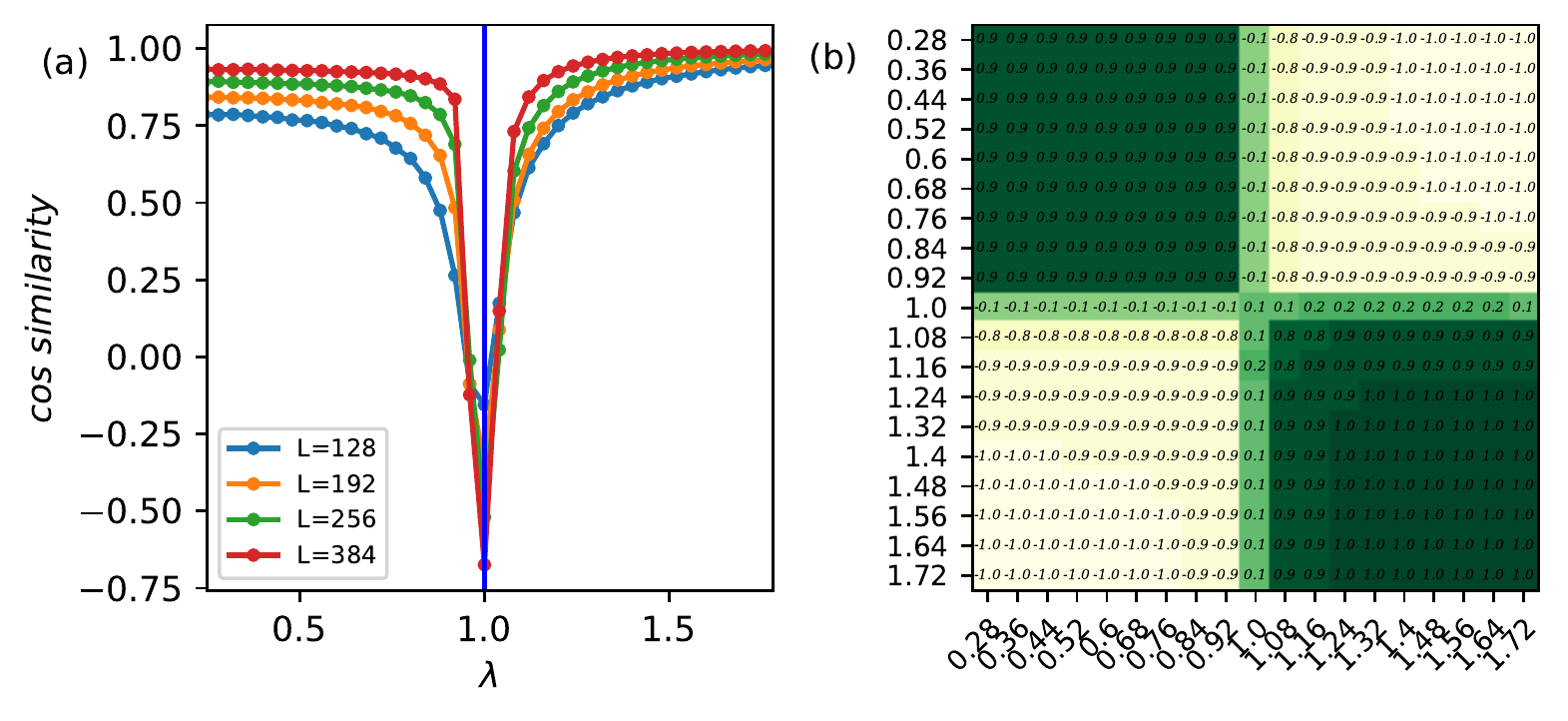}
  \caption{Cosine similarities estimated for the Aubry-Andr\'e model of spinless fermions. (a) Adjacent similarity, which measures the similarity between state configurations of the single-particle ground state with disorder strength's $\lambda - \Delta \lambda$ and $\lambda + \Delta \lambda$. System sizes $L = 128, 192, 256$, and $384$. (b) Mutual similarity with $L = 384$, which measures the similarity between state configurations with any two disorder strength's $\lambda$ and $\lambda^{\prime}$.}
  \label{fig:aamodel}
\end{figure}

As shown in Fig.~\ref{fig:aamodel}, similar to the Ising model and the quantum compass model cases above, the adjacent similarity is close to 1 (saturation value) for $\lambda$'s far away from $\lambda_c = 1$ and presents a V-shape around $\lambda \sim \lambda_c$. Our method successfully predicts the phase transition point and works also for this quantum model with a disorder-driven quantum phase transition.

\subsection{Contrastive learning phases of the Su-Schrieffer-Heeger model}

As the last example, we study the Su-Schrieffer-Heeger (SSH) model \cite{Heeger1988SSH}, which features a one-dimensional topological insulator characterized by a global topological invariant---the winding number, and protected by chiral symmetry. \cite{shen2017topological} The Hamiltonian reads
\begin{align}
  H =& -(J + \kappa) \sum\limits_{i}{(c_{i,A}^{+}{{c}_{i,B}} + h.c.)} \nonumber \\
     & -(J - \kappa) \sum\limits_{i}{(c_{i,B}^{+}{{c}_{i+1,A}} + h.c.)},
  \label{eq:SSH}
\end{align}
where $c_{i,A(B)}^+$ and $c_{i,A(B)}$ are the creation and annihilation operators of the spinless fermions on the $A$($B$) sublattice site in the $i$th unit cell, respectively. Each unit cell consists of two sites, $(i,A)$ and $(i,B)$. The hopping terms always connect two adjacent lattice sites, an $A$ sublatice site and a $B$ sublattice site. The hopping amplitude in the unit cell is $-(J + \kappa)$, and that between two neighboring unit cells is $-(J - \kappa)$. We set $J = 1$ as the energy unit. As is well known, the half-filled system is a trivial insulator when $\kappa > 0$ and is a topological insulator when $\kappa < 0$. A topological phase transition occurs at $\kappa = 0$. Standard variational Monte Carlo provides us with state configurations weighted by the amplitude squared of the ground state wave function in the real space. The training and testing data sets are prepared for $N = 51$ $\kappa$'s, $\kappa_k = -0.75 + (k - 1) \Delta \kappa$ and $\Delta \kappa = 0.03$. The two-layer MLP is used as the encoder neural network, the same as the Ising model case above.

\begin{figure}[htp!]
  \includegraphics[width=8cm]{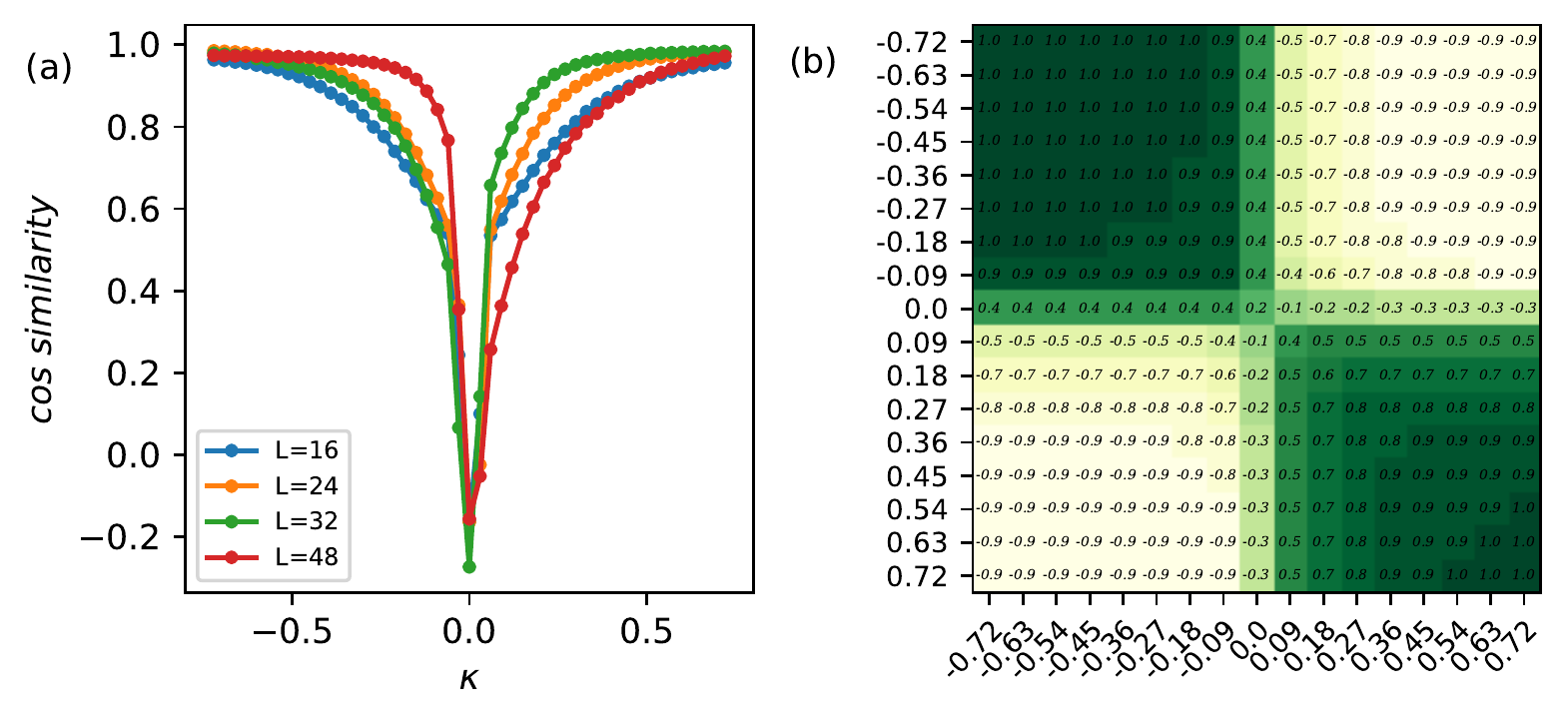}
  \caption{Cosine similarities estimated for with the SSH model. (a) Adjacent similarity, which measures the similarity between state configurations of the ground state with $\kappa - \Delta \kappa$ and $\kappa + \Delta \kappa$. System sizes $L = 16, 24, 32$, and $48$. (b) Mutual similarity with $L = 48$, which measures the similarity between state configurations with any two model parameters $\kappa$ and $\kappa^{\prime}$.}
  \label{fig:SSH}
\end{figure}

As shown in Fig.~\ref{fig:SSH}(a), similar to the previous models, the adjacent similarity is close to 1 (saturation value) for $\kappa$'s far away from $\kappa_c = 0$ and presents a V-shape around $\kappa \sim \kappa_c$. Our method successfully predicts the topological phase transition point.

From a conventional (local) point of view, the topologically insulating phase ($\kappa < 0$) and the trivially insulating phase ($\kappa > 0$) are difficult to discriminate because the two phases are similar locally. On the other hand, they can be evidently discrminated through the topological invariant, i.e., they are different globally. Remarkably, as shown in Fig.~\ref{fig:SSH}(b), {SimCLP} can readily discriminate the two topologically different phases. The mutual simmilarity is close to 1 (maximal simmilarity) for state configurations in the same phases and is close to $-1$ (maximal difference) for state configurations from the two different phases. It is worth emphasizing that {SimCLP} has done this only by learning from state configurations, without complicated mathematics or phase-specific tricks in other methods, showing the simplicity and broad applicability of our framework.

\section{Summary and outlook}

In summary, we have proposed a framework for contrastive learning phases of matter and shown that it can encode phases of various systems, no matter whether classical or quantum, single-particle or many-body, conventional or topological, and can predict phase transitions readily when enough state configurations are provided. Our framework is simple and flexible. It does not need any prior knowledge such as data labels and order parameters. As valuable by-products, it can provide representation vectors and labels for state configurations, which may be used for other purposes, such as ground state representation, \cite{Carleo2017Science,Glasser2018NNQS,Cai2018wavefunctions} accurate determination of phase transition points and critical exponents, \cite{sancho2021kernels,Zhang2019percolation,Bachtis2020symmetry} quantum error correction protocols, \cite{Torlai2017Topological} and quantum state tomography \cite{torlai2018tomography}.

\begin{acknowledgments}
This work was supported by National Natural Science Foundation of China (Grants No. 11874421 and No. 11934020) and Beijing Natural Science Foundation (Grants No. Z180013). Computational resources were provided by Physical Laboratory of High Performance Computing in RUC.
\end{acknowledgments}

\bibliography{simclp}

\end{document}